# Approximate ADCs for In-Memory Computing

Arkapravo Ghosh[1], Hemkar Reddy Sadana[1], Mukut Debnath[1], Panthadip Maji[1], Shubham Negi[2], Sumeet Gupta[2], Mrigank Sharad[1], Kaushik Roy[2]
[1]Indian Institute of Technology Kharagpur, Purdue University

*Abstract*—In memory computing (IMC) architectures for deep learning (DL) accelerators leverage energy-efficient and highly parallel matrix vector multiplication (MVM) operations, implemented directly in memory arrays. Such IMC designs have been explored based on CMOS as well as emerging non-volatile memory (NVM) technologies like RRAM. IMC architectures generally involve a large number of cores consisting of memory arrays, storing the trained weights of the DL model. Peripheral units like DACs and ADCs are also used for applying inputs and reading out the output values. Recently reported designs reveal that the ADCs required for reading out the MVM results, consume more than 85% of the total compute power and also dominate the area, thereby eschewing the benefits of the IMC scheme. Mitigation of imperfections in the ADCs, namely, non-linearity and variations, incur significant design overheads, due to dedicated calibration units. In this work we present peripheral aware design of IMC cores, to mitigate such overheads. It involves incorporating the non-idealities of ADCs in the training of the DL models, along with that of the memory units. The proposed approach applies equally well to both current mode as well as charge mode MVM operations demonstrated in recent years., and can significantly simplify the design of mixed-signal IMC units.

## I. INTRODUCTION

Deep Neural Networks (DNN) involve a large number of dot-product calculations between multiple pre-trained convolution kernels and input pixels at each layer [1, 2]. Such kernel multiplications over pixels from multiple input-channels can be organized in the form of matrix-vector-multiplication (MVM) operations. Conventional DNN accelerator architectures incur significant performance and power overhead due to extensive data movement between multiple processing units and associated local and shared memory blocks, storing these kernel weights and partial results [3, 4].

In-memory computing (IMC) scheme has gained significant attention for energy-efficient DNN accelerator design, owing to reduced memory-to-processor data-traffic [5-7]. IMC involves storing kernel weights for each layer and its input channels, in memory arrays. The input data can be applied along the word-lines. Based on the circuit level interaction between the applied signal level and memory cell, current-mode [11-13] or charge-mode [14, 15] dot product is computed. The result of the dot product operation is generated along the column-wise bit-lines of the memory array, which accumulate the current or charge outputs from multiple memory cells. The resulting outputs are inevitably analog values, which need to be quantized into digital outputs, before feeding them to the next layer. The number of quantization levels depend upon several parameters like, number of input levels, number of weight-bits and size of the memory-array. Larger values of each of these parameters translate to a larger number of possible output levels and hence mandates larger quantization levels for overall higher inference accuracy [37]. Larger levels of output quantization require higher resolution analog to digital converters (ADC) for digitizing the result [26, 27]. The ADC resolution requirements as a function of the aforementioned design-parameters has been estimated in recent work [38, 39]. The power and area overhead associated with ADCs grows sharply with increasing bit resolution, while the performance drops [38]. Recent designs ascribe more than 85% power consumption to the ADCs and only a smaller fraction is used for actual analog-mode IMC operation in the compute core [40, 41].

For a target bit resolution, design for higher linearity, measured in terms of INL and DNL (integrated and differential non-linearity), and reduced variation-effects, results in higher power, area and performance penalties [16-18]. Ideally, for highest possible performance from IMC cores, the number of ADCs should be equal to the number of memory array columns. However, large area foot print of ADCs necessitates their sharing among multiple columns.

To overcome the challenges related to ADCs in IMC units, we propose ADC-aware DNN training method, which involves the use of non-ideal and imperfect ADC characteristics with variations and non-linearity, in place of ideal ReLU functions of the convolution layers (fig. 1). The DNN model is first trained with ideal ReLU (perfectly linear and variation free) and weight values, with bit-truncation, to estimate the minimum bits required to retain the inference accuracy close to that obtained through high-resolution floating-point ReLU and weights. The non-linearity and variations estimates obtained from un-calibrated ADC of required bit-resolution, estimated from circuit design, is then incorporated in the model re-training, by replacing the ReLU with such imperfect characteristics. The ADC design constraints are tightened till the re-training achieves a desired accuracy, close to the ideal ReLU case. The tolerance towards weight variations and subsequent bit-cell sizing constraints are obtained in parallel, through addition of estimated weight-noise from bit-cell and array simulations. VAT approach for weight variations has been explored earlier [47]. Incorporation of ADC imperfection in VAT, proposed in this work can significantly simplify the associated design constraints for low power and high-performance IMC architectures.

Rest of the paper is organized as follows. A brief description of IMC unit along with the peripheral circuits in given in section-II. Section-III presents design analysis and characterization of an ADC based on ring oscillators. Variation aware training framework is described in section IV. Section V presents comparison with conventional

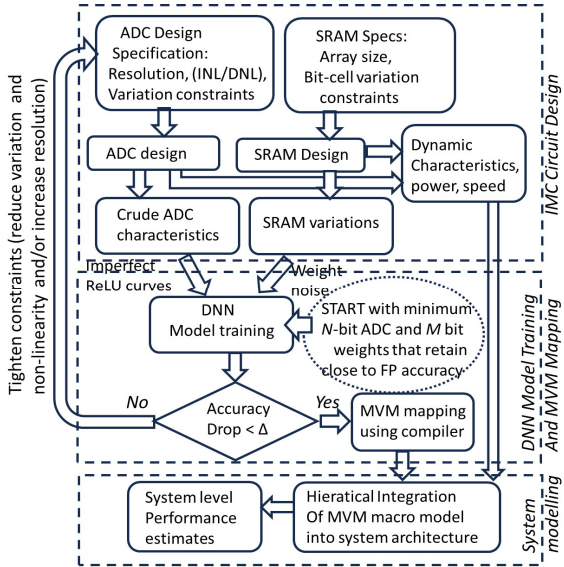

Fig. 1a. Mapping of DNN to an NVM IMC unit

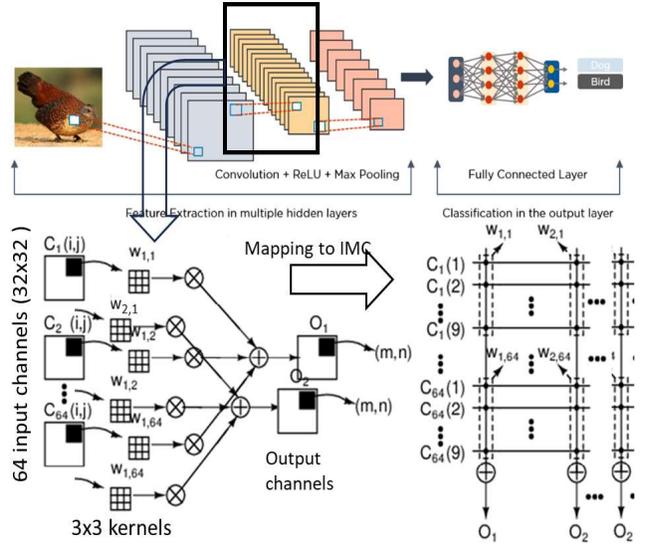

Fig. 1b IMC mapping of CNN

approach of calibration. Conclusions are given in section VI.

## II. IN MEMORY COMPUTING MACRO

In this section we present a brief description of IMC scheme based on SRAM . Both, current-mode [11, 19, 20], as well as, charge mode MVM computation [14, 15, 21] based on SRAM are discussed.

### A. Mapping of DNN onto IMC SRAM Macro with ADC Based Readout

As shown in fig.1b, the convolution operation involved in a DNN architecture can be represented as MVM operations [22, 23]. To obtain the output pixel of a particular layer, convolution of all kernels of that layer with the pixels of all input channels of the previous layers are needed. A fully parallel implementation of such a compute step would require large IMC crossbars of dimension $M \times N \times S^2 \times P$, where $M$ is the number of input channels to the layer, $N$ is the number of kernels, $S$ is the size of kernels and $P$ is the number of output channels, assuming that all output pixels of a particular layer are computed in the same IMC array. An IMC core, in general constitutes of a memory array (SRAM/DRAM/NVM) along with input and output- readout peripheral circuits. For mixed-signal implementations, digital inputs to a particular IMC core are applied to row-parallel DAC, as shown in fig.1. The DAC outputs drive the word-lines. Each column accumulates the result of dot products between the input values and the kernel weights stored in the IMC memory elements in the respective columns. The analog-mode output thus produced, needs to be digitized by ADCs, before getting applied to the next layer operation. In terms of degree of parallelism and hence, performance, larger array-size and higher bit resolution for inputs, weights as well as outputs are favorable. However, larger array size and bit resolutions of weights translate to requirement of higher resolution ADCs [24-27].

This is because, the ADC needs to resolve more levels in the analog-mode MVM output for such cases. High precision ADCs can overwhelm area and power cost and hence are not suitable for high-performance column-parallel operations. Use of larger IMC arrays also suffer from degradation in compute precision due to the effect of parasitic and device variations [8, 33, 34]. The MVM operation therefore needs to be partitioned and allocated to smaller memory arrays.

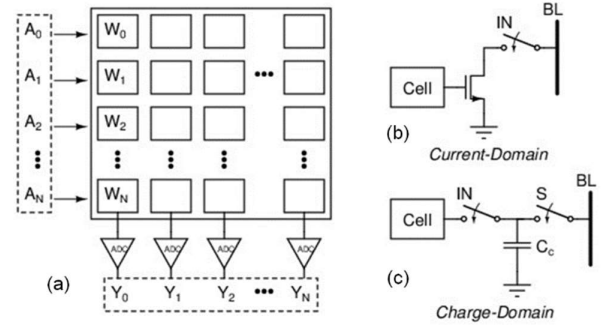

Fig. 2 (a) IMC unit based on SRAM, with (b) current-based and (c) charge-based bit cell operation for MVM

Bit-wise computation schemes for IMC cores have been proposed to reduce the requirements for ADC resolutions [24-27]. This is achieved by allocating different memory arrays for each of the weight-bits, feeding inputs in a bit-wise manner and combining the partial sums obtained from each of the arrays, using digital peripherals through shift-add operations [28-29]. While, the use of multi-level RRAM cells have been proposed for implementing multi-bit weights in a single array [42], such weight-bit slicing facilitates IMC implementation using 1-bit SRAM memory cells, as shown in Fig.2a. Streaming inputs in bit-wise manner also eliminates the need for multi-bit DACs at the row-input points. Even for bit-wise operations, a moderate size memory array of dimension 64x64, leads to 6-to-8-bit resolution requirement for ADCs [9, 35, 36], as shown in a subsequent section. Partial word-line activation has also been used to compute lower-bit partial-sums corresponding to smaller sets of inputs applied sequentially [30-32]. However, such approaches transfer significant amount of computation to digital peripheral circuits which combine the partial results obtained from multiple cross-bars. This results in increase in area and power overhead and also incurs performance penalty due to reduced degree of parallelism.

*B. Current-mode IMC Operation*

For SRAM based IMC cores, two different modes of MVM operation have been proposed in recent years, namely: current mode [11, 19, 20] and charge-mode [14, 15, 21]. As shown in Fig. 2b, current mode operation for SRAM core, involves conditionally drawing a static current from the bit-line, in the weight-bit (stored in the SRAM bit cell) and the input-bit are both high. The net current flowing through the bit-line represents the dot-product between the weight bits stored along the column and the input bit-vector. As shown in Fig. 2c, the charge-mode scheme involves an additional capacitor in each bit-cell to conditionally store and transfer charge to the bit-line, depending upon the weight and input bit values [15, 21]. It essentially involves charge-sharing between the individual bit-cell caps and the bit-line cap, to implement the summation of bit-wise dot products obtained through individual cells along the column. Though charge-mode operation offers some benefits, like reduced static power and faster compute speed and lesser variation, it may suffer from poorer scalability due to added capacitors per bit cell [15, 43]. The VAT scheme proposed in this work is applicable to both cases. However, we limit our discussion to current-mode computation in this work.

*C. ADC for IMC operation*

For current-mode IMC, the transimpedance stage can be used to convert the output current into a proportional voltage. Alternatively, a current integrator stage can also be used to produce a proportional output voltage [44]. The resulting voltage signal can be applied to a voltage mode ADC. Single slope ADC (SS-ADC) [46], as well as Successive Approximation Register ADC (SAR-ADC) [45, 46], have been used for current-mode IMCs. While SS-ADC are relatively compact, they need large number of clock-cycles per conversion ($2^N$ cycles for $N$ bit output). SAR ADCs incur significant area overhead due to the capacitive DAC, but take only $N$ clock cycles for a single $N$ bit conversion.

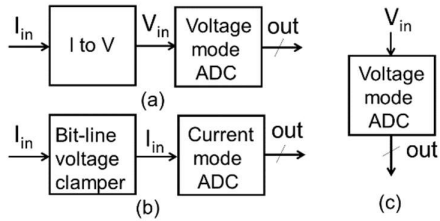

Fig. 3 ADC interface for current-mode IMC column using (a) TIA and voltage-mode ADC, (b) current-mode ADC, and (c) for charge-mode IMC using voltage mode ADC

An alternate approach for current mode IMC is the use of current-mode ADCs, which can help eliminate transimpedance operation and the associated circuit overhead (fig. 3b). For instance, a current-controlled oscillator can be used to generate oscillations with frequency proportional to the input current, which can be converted into the output digital code with the help of a counter [11]. Though the current-mode ADC overcomes the need of the transimpedance stage, it does require a biasing stage at the input, which clamps the bit-line voltage to a desired level. In this work we present detailed analysis of current-mode ADC, however, similar benefits are expected from voltage mode ADCs as well.

### III. CCO-BASED ADC FOR IMC MACRO

The current-mode ADC used in this work consists of a ring-oscillator based current-controlled oscillator (CCO) [11]. Due to it's simple and compact structure, it can be amenable to column-parallel operation, i.e., each column of the SRAM IMC array having a dedicated ADC for highest possible read-out speed. CCO oscillates at a frequency which has a near-linear dependence on the input current $I_{BL}$, received from the bit-line (Fig. 4). Following the CCO is a ripple-counter which produces the final digital code at the end of the evaluation period $T_{eval}$. For interfacing the CCO with the bit-line, an OPAMP-based negative feedback loop is used to maintain a constant bit-line voltage, ensuring a fixed voltage drop $V_{BL} = V_{ref}$, across the current sinking device of all bit-cells, as shown in fig. 2b. For nominal supply voltage, the frequency vs input current characteristics are apparently linear. The influence of supply-voltage scaling on the frequency versus control current characteristics of the CCO in nominal (TT) corner is presented in Fig. 5(a).

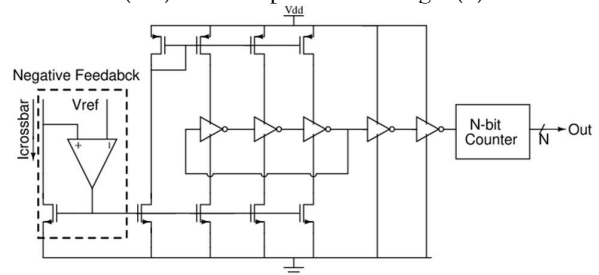

Fig. 4. Block-diagram of CCO - based ADC

We choose the maximum CCO frequency ($f_{max}$) and thus, the conversion speed based on the maximum $I_{BL}$ ($I_{max}$) when all the dot-products are HIGH in a column. This can be simplified for an $N$-bit (=7) ADC as shown in (1).

$$t_{eval} = \frac{2^N}{f_{max}} = \frac{2^N}{k_{CCO} \cdot I_{max}} \quad (1)$$

The ADC output as a function of the column current ($I_{BL}$) is shown in (2).

$$CODE = t_{eval} \cdot f_{CCO} = \frac{2^N \cdot I_{BL}}{I_{max}} \quad (2)$$

The average ADC computation power and energy is found to reduce with the supply voltage (fig. 5b). This is because the input current, which determines the oscillation frequency is independent of the $V_{DD}$. Hence, supply voltage scaling is advantageous in terms of power and energy efficiency (fig. 5c). However, linearity of the CCO and hence the ADC drops

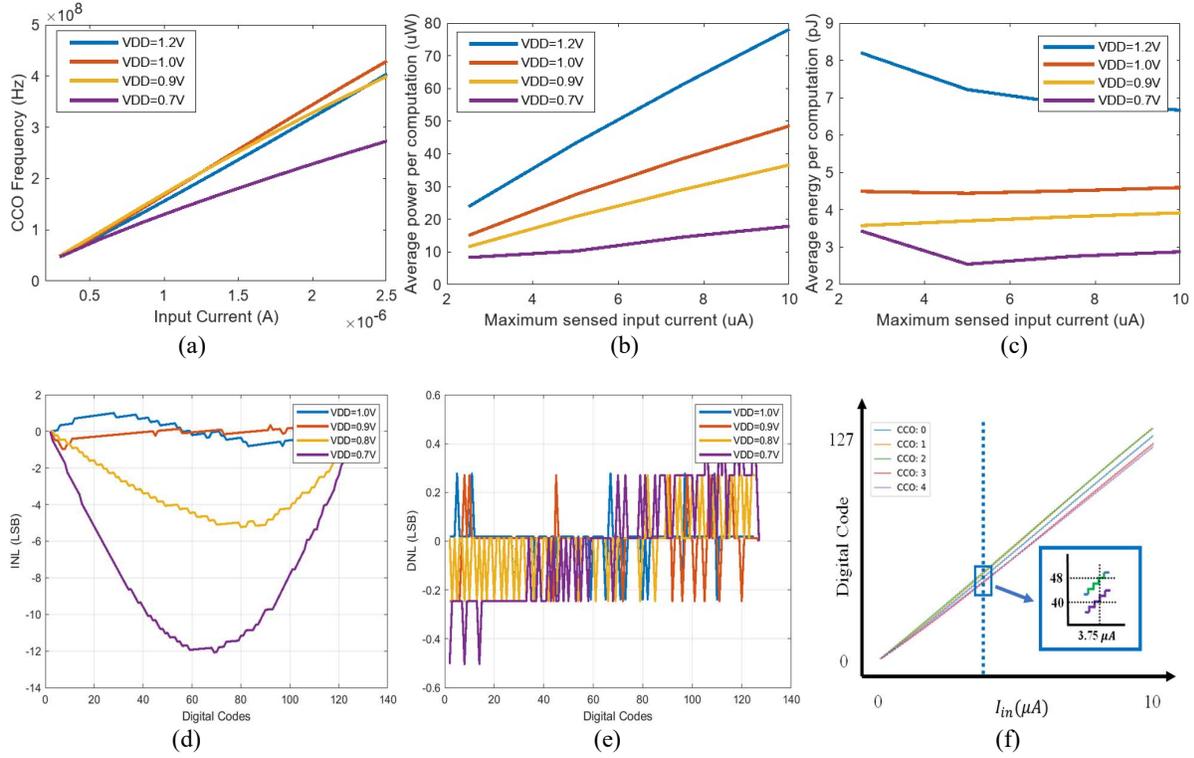

Fig. 5. Trends in performance metrics of CCO-ADC: (a) CCO frequency characteristics (b) ADC compute power v/s maximum BL current (c) ADC compute energy v/s maximum BL current (d) INL characteristics for $I_{max} = 10\ \mu A$ (e) DNL characteristics for $I_{max} = 10\ \mu A$, (f) ADC characteristics across worst-case process corners

with reducing supply voltage. This is due to increasingly imperfect mirroring of input current into the CCO unit from OPAMP controlled input branch, resulting from the reduced voltage headroom for the current mirrors. This is evident from the INL and DNL plots of the ADC shown in fig. 5d and fig. 5e. In general, similar trend of power consumption and linearity is observed in other ADC topologies.

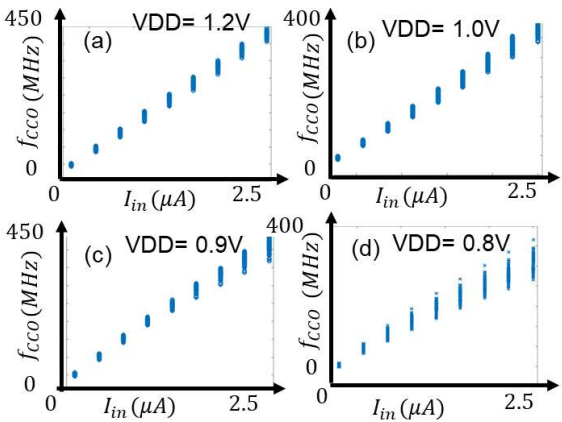

Fig. 6. Effect of random variations on CCO frequency characteristics for different supply voltages: showing higher spread and non-linearity for lower supply.

Change in the slope of the characteristics across corners, as shown in fig. 5f, can be ascribed to the shift in device threshold-voltage mean-values and hence the stage delays of the CCO. Such corner-wise PVT-shifts, which would affect all ADCs on a die in a similar fashion, can be compensated through global calibration, which involves corner detection circuits and applying corner dependent scaling of digitized outputs [49].

Random variations on the same die can cause significant differences between ADC characteristics on the same chip. The effect of random variations obtained through Monte Carlo simulations is shown in fig.6, for different supply voltages. It indicates that the spread in the CCO characteristics due to random variations increases with the down-scaling of the supply voltage. This is due to the devices being increasingly pushed towards near-threshold operation, resulting in reduced current density and higher percentage variations. Hence, both variation and power consumption trade-off with energy-efficiency. In general, mitigating such random variations would require dedicated calibration of each ADC, which can incur significant design, area and performance overhead [11].

## IV. VARIATION AWARE TRAINING FRAMEWORK

In this section, we describe the VAT scheme which addresses the non-idealities of ADCs along with the SRAM and crossbar interconnects.

### A. Modeling of crossbar with non-idealities

The 8T SRAM unit cell, depicted in Fig. 7, has the output port constituted by M7, driven by the input word-line and M8, driven by the bit-cell weight value (high or low). The current sunk by M7-M8 in a sizable SRAM array can suffer from inaccuracy due to parasitic resistances arising from the read-bit-line (RBL) interconnect. The dynamic characteristics of the output signal is affected by parasitic caps of the transistors as well as the RBL. The parasitic resistance and capacitance per unit length of the word-line (WL) are designated as $r_{wl}$ and $c_{wl}$, while the bitline (BL) parasitics are represented by $r_{bl}$ and $c_{bl}$. For a given technology node, these parameters equate to the wire resistance ($r_w$) and the wire capacitance ($c_w$) per unit length. The total parasitic resistance and capacitance between two unit cells are expressed $as\ R = r.L$ and $C = c.L$, where $r = r_{bl}$ or $r_{wl}$, and $c = c_{bl}$ or $c_{wl}$. Notably, the effect of the parasitics become more pronounces for increasingly larger array sizes. The effect of these parameters have been incorporated in the training model using the open-source GENIEx framework [49]. GENIEx stands for Generalized Approach to Emulating Non-Ideality in memory arrays by using Neural Networks (NN). Using a trained NN-based, it estimates the output signal of the RBL based on conductance values of the bit-cell elements ( in this case, M7-M8), parasitic and the applied input.

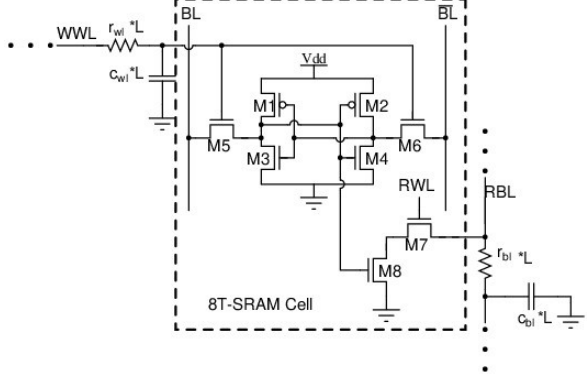

Fig. 7. Modeling of interconnect parasitics in crossbar

*B. Description of Training process used:*

In this work, we train the ResNet20 model on CIFAR10 datase. This model comprises 19 convolution layers along with a fully-connected classifying layer at the output. Employing the optimal 32-bit floating point configuration within the ResNet20 PyTorch architecture [], a classification accuracy of 93.26% was obtained. Subsequently, the minimum weight-bit precision and the ideal ADC resolution required to retain an inference accuracy close to the ideal case was estimated using iterative training.

**(i)Pytorch training of selected CNN model with target bit precision of input and weights.**

Fig. 8 depicts a boxplot, excluding outliers, illustrating the distribution of weights and Conv2d outputs under ideal 32-bit floating-point precision arithmetic. Notably, the weights predominantly fall within the range of 0 to 1, while the Conv2d layer outputs span from 0 to 4. Based on this observation, 0 integer bits for weights and 2 integer bits for ADC were allocated and number of fractional bits were estimated by iteratively computing the training accuracy , as shown in table-I

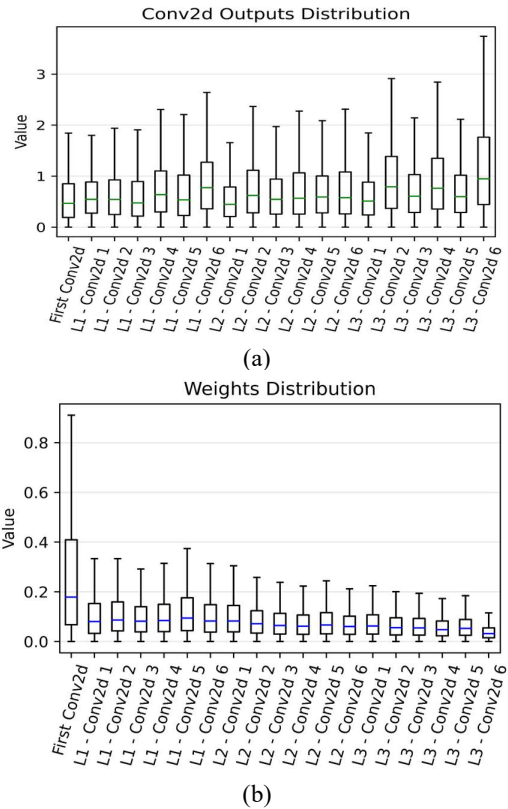

Fig. 8. Distribution of the pretrained ResNet20 model for (a) Conv2d (ReLU) outputs (b) Conv2d Weights

.Table-I
Training accuracies for 0 integer bit for weights and 2 bits for activations

| Weight Bits | Activation Bits | | | |
|---|---|---|---|---|
| | 5 | 6 | 7 | 8 |
| 6 | 40.33 | 84.62 | 91.06 | 92.21 |
| 7 | 59 | 81.34 | 92.05 | 92.32 |
| 8 | 50.42 | 80.79 | 92.96 | 92.14 |
| 9 | 41.56 | 81.28 | 92.97 | 92.36 |

The table presents training accuracies corresponding to various total numbers of bits, while maintaining fixed 0 integer bits for weights and 2 integer bits for activations, as dictated by necessity. Notably, the findings reveal that employing a total of 7 bits for both weights and activations yields only a marginal decrease in accuracy compared to the accuracy achieved with the ideal floating-point model. Consequently, we opt to proceed with utilizing 7 bits exclusively for subsequent training and testing phases.

**(ii) Retraining with real ADC characteristics to mitigate effects of non-linearity.**

We first test the impact of non-ideal ADC characteristics on the inference accuracy. Here, ideal ADC is referred to the 32-bit floating-point (FP) ReLU transfer-function, which acts as a perfectly linear ADC with no variations. First, the model trained on ideal ADC characteristics is tested with non-ideal ADCs, results for which are shown in table-I. Since, 7 bit ideal ADC characteristics was found to produce inference accuracy close to the 32-bit case, for subsequent analysis 7-bit ADCs were used. The first case is ideal 7-bit ADC, which is obtained by replacing the 32-bit ReLU function by a 7-bit counterpart. We also test the model accuracy with 7-bit CCO ADC characteristics obtained in the typical corner. For comparison, we also used a single-slope ADC characteristics. The 2$^{nd}$ column shows degradation in inference accuracy due to incorporation on non-ideal ADC characteristics. This is due to non-zero INL and DNL, for such ADCs (as observed earlier in case of CCO ADC).

Next, retraining (RT) of the ResNet20 model was performed with non-ideal ADC characteristics. This was done by replacing the ideal ReLU operation with simulated ADC characteristics for a particular corner. Subsequently, the model thus trained was tested with the same ADC characteristics to evaluate the impact of retraining. The 3$^{rd}$ column in table-I shows that, post retraining, the inference accuracy was restored for the real ADC characteristics. This shows that the retraining process makes the model robust towards the non-linearity and offsets present in real ADC characteristics.

Table-II
Test Results of different ADCs

| ADC Scheme | Test Accuracy | |
|---|---|---|
| | w/o RT | with RT |
| Ideal (32 bit FP ReLU) | 93.26 | 93.26 |
| Ideal (7bit) | 92.73 | 93.05 |
| SAR ADC (7bit) | 84.69 | 93.06 |
| CCO ADC (7bit) | 75.13 | 92.81 |

**(iii) Variation Aware Training for ADC**

Upon establishing the efficacy of retraining with real ADC characteristics to maintain classification accuracy, we proceeded to assess the trained model's performance under random variations, captured by Monte Carlo (MC) simulations. 200-run MC samples were used to model the variation of column-parallel ADC for 128x128 IMC array. First, one randomly selected sample allocated for training and the rest utilized for testing to evaluate the impact of random variations in ADCs on inference accuracy. Our findings reveal a substantial drop in accuracy due to these variations across large number of tests, as shown in fig. 9a.

To address this, we introduce Variation Aware Training (VAT), wherein all 200 MC characteristic samples are incorporated during the training process, randomly allocated to each output node of a neuron. This approach ensures the training process captures the circuit variations derived from Monte Carlo sampling and restores the model accuracy close to the mean-value of inference accuracy for the case without variation, as shown in fig. 9b. Consequently, the proposed approach achieves significant reduction in accuracy-drop due to random variations in ADCs, as the neural network learns to adapt to such characteristics.

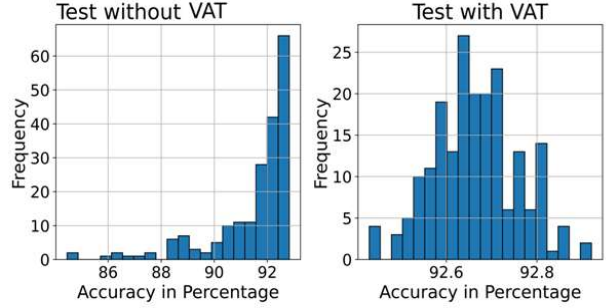

Fig. 9. Test results on CCO ADC variations

(iv) Modelling of weight variation

Finally, we use the model proposed by [47], for the modeling of the weight variations. These variations are incorporated during the training process itself. The noise is added to the 7-bit quantized weights of the Conv2d layers of the CNN model just before the application of the forward pass in each iteration. First, for a given $\gamma$, we analyze the CNN performance under the following conditions: (i) non-noisy weight and ideal ADC (uniform quantization), (ii) noisy weight and ideal ADC, (iii) noisy weight and single CCO ADC, (iv) noisy weight and CCO ADC MC VAT. Note than $\gamma = 0.1$ effectively translates to $\sigma/\mu$ of 10%, where $\sigma$ is the standard deviation and $\mu$ is the mean of the effective weight value. This directly translates to the distribution of bit-cells compute current sunk by the output device, as in fig. 2a.

Table-III
Incorporating weight noise in training

| Condition | Accuracy (%) |
|---|---|
| Ideal ADC, $\gamma = 0.0$ | 93.05 |
| Ideal ADC, $\gamma = 0.1$ | 91.84 |
| CCO ADC, $\gamma = 0.1$ | 91.50 |
| CCO ADC VAT, $\gamma = 0.1$ | 90.38 |

Table-III shows that weight variations has significant impact on accuracy post-training. The results may be improved by upsizing of the SRAM output device, to reduce the spread in output current and hence the effective weight, specifically, for the SRAM arrays computing for higher value weight-bits (increasing the number of weight bits. It's important to note that, only addressing the weight-bit variation and ignoring the ADC non-idealities would lead us back to results shown in table-II, without retraining.

### V. COMPARISON WITH CALIBRATED ADC

In this section we present describe a more conventional scheme of single point calibration for multiple ADCs associated with an SRAM based MVM unit for mitigating the effects of random variations. We compare this approach with the proposed VAT scheme. We also present the variation results for another popular ADC topology for IMC, namely SAR-ADC, to justify the need of the proposed scheme IMC design, irrespective of the ADC topology.

### A. Comparison with Single Point Calibration Scheme.

As compared to more complex calibration scheme used in [11] for CCO based ADC, we adopt a relatively simpler, single-point calibration scheme in this work. This scheme forces the different ADC characteristics to coincide at the point of calibration. We observe that the random variations effect the slope of the CCO characteristics, (fig. 5f and fig. 6). Assuming that the slope remains roughly constant (which is a significant simplification for low voltage operation), we provide a correction factor for it by tuning the current-mirroring ratio between the input branch of the CCO and the current source devices of the starved inverters. The detailed circuit-level implementation of the proposed single-point calibration scheme is shown in Fig. 10. The combinational logic unit produces a 9-bit digital output that serves to selectively enable or disable a bank of equally sized transistors connected in parallel, within the tuning bit circuit. This method of tuning is referred to as the thermometer coding scheme.

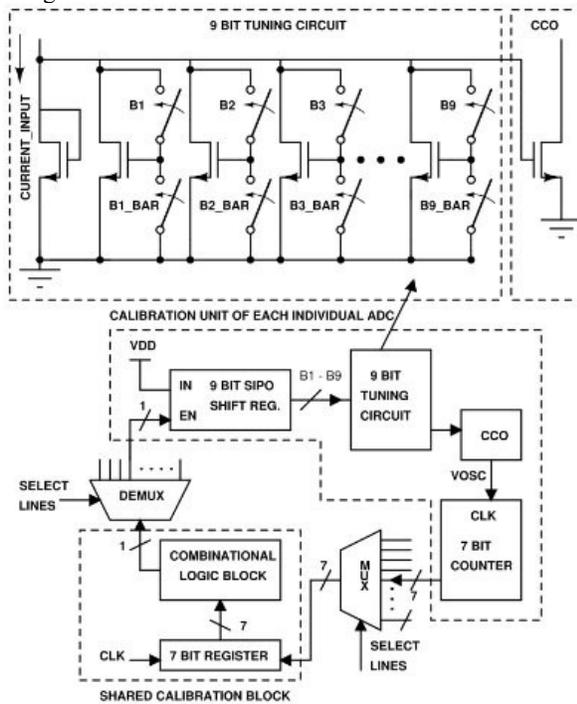

Fig. 10. Single point calibration scheme for CCO ADC.

The calibration starts with resetting the serial-in-parallel-out (SIPO) shift register. At every clock cycle, one bit of the shift register turns HIGH which turns ON one MOSFET in the tuning block. The moment at which the digital output becomes less than our reference count, the combinational logic circuit disables the register, thus fixing its content. The logic circuit of the calibration unit can be shared across multiple ADCs of different columns using a proper bus connection to calibrate the ADCs sequentially. However, since each ADC can have different tuning bits owing to random mismatch, the tuning block needs to be separate.

The ADC characteristics post calibration are shown in Fig. 11. It shows significant reduction in spread as compared to the plot in fig. 5f. Here, the calibration point is chosen around $1/3^{rd}$ of the maximum targeted current. It can be observed that the calibrated characteristics tend to diverge towards higher current values while they are well matched for current values up to middle of the range. This is purposefully done, considering the output statistics of each convolution layer. More than 75% of the outputs are found to be less than 25% of the maximum value for all layers and hence, matching the ADCs better at lower values is advantageous.

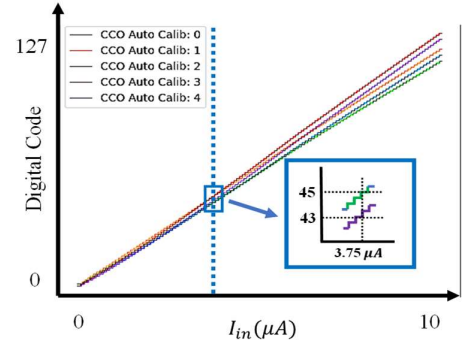

Fig. 11. ADC characteristics across process corners post-calibration

For comparing the area penalty of this calibration scheme, the layouts of 7b CCO-ADC with and without calibration block are shown in Fig. 12. The proposed calibration scheme results in an area penalty of around 3x, due to additional components like switches, registers and control signals.
The training accuracy post calibration was found to be around 88.5% , which was lower than the case of variation aware training. It may be possible to improve the training accuracy by employing more sophisticated calibration schemes as in [11], however, that would come at a further higher area penalty.

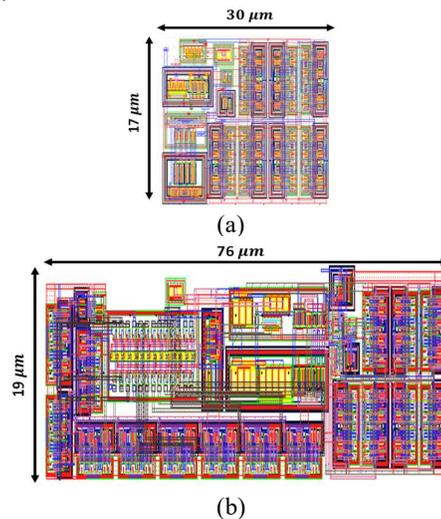

Fig. 12. 7b CCO-ADC layouts (a) with and (b) without calibration

## B. Variations in SAR ADC and need for VAT.

In order to interface a SAR ADC with an SRAM based IMC array, with current mode operation, we need to convert the output current into proportional voltage. This can be achieved using a transimpedance amplifier or an integrator as shown in fig. 13. Hence, we need to incorporate the effects of variations and non-linearities in both, the ADC as well as integrator.

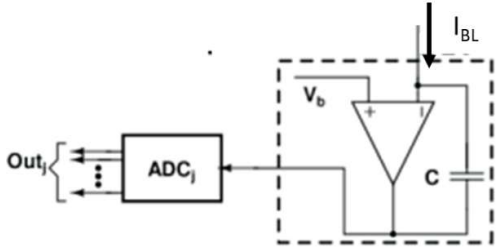

Fig. 13 SAR ADC interface with IMC output.

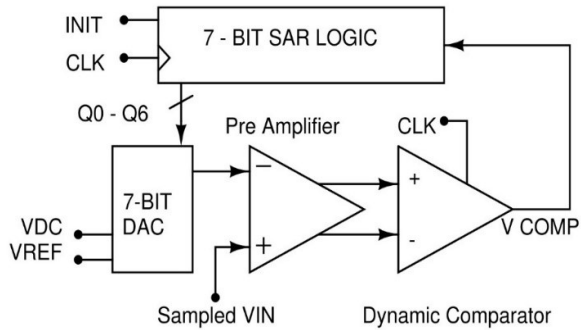

Fig.14 Schematic diagram of SAR ADC.

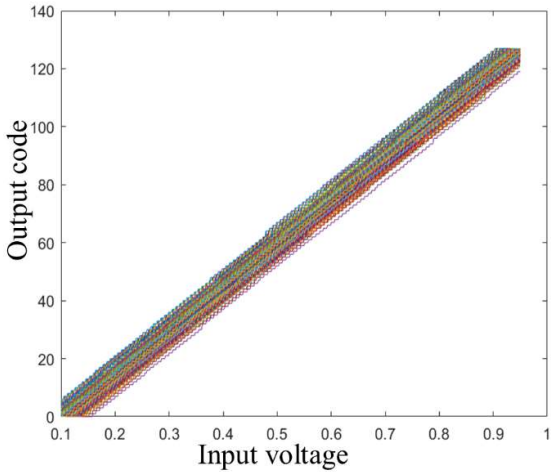

Fig. 15 MC results for 7-bit SAR characteristics without any calibration .

SAR ADC constitutes of variation sensitive units like the capacitive digital to analog converter (DAC) and the comparator. Both can contribute to significant variations resulting from capacitive mismatch and offset respectively. Fig. 15 shows the 200 run MC simulation results for a 7 bit SAR ADC in 65nm CMOS technology. The efficacy of the proposed VAT scheme was tested for the SAR ADC characteristics as well. Without VAT, the model accuracy was less than 81%, which improved to 91.2% after the variation aware training process. This approach can therefore allow smaller unit-caps and smaller devices for the comparator and integrating amplifier, as it can allow the IMC core to work with cruder and less precise ADCs, without the need of per-device calibration. Smaller devices can facilitate lower area, higher speed as well as lower power operation for large number of ADCs interfaced with the IMC units.

## VI. CONCLUSION

In this work we proposed variation aware training approach for mitigating the impact of ADC non-idealities for In-Memory-Computing. Conventional approaches involve expensive calibration scheme and mandate relatively robust and accurate ADC designs for integration with IMC cores. The proposed approach on the other hand can significantly relax the design constraints for the read-out ADCs and allow relatively more compact units. The VAT scheme proposed in this work can be extended to different IMC schemes, including charge-mode and current-mode methods. It is also applicable to different ADC topologies suitable for IMC integration. The proposed approach can be integrated into system level design flow for DNN accelerators based on IMC, which would result in circuit-architecture-algorithm co-design.